\documentclass[12 pt, amsfonts,amsmath, amssymb,color]{article}

\usepackage{amsmath,amssymb,amsfonts,latexsym,float,graphics,epsfig}

\begin{document}

\renewcommand\theequation{\arabic{section}.\arabic{equation}}
\catcode`@=11 \@addtoreset{equation}{section}
\newtheorem{axiom}{Definition}[section]
\newtheorem{theorem}{Theorem}[section]
\newtheorem{axiom2}{Example}[section]
\newtheorem{lem}{Lemma}[section]
\newtheorem{prop}{Proposition}[section]
\newtheorem{cor}{Corollary}[section]
\newcommand{\be}{\begin{equation}}
\newcommand{\ee}{\end{equation}}
\newcommand{\ben}{\begin{equation*}}
\newcommand{\een}{\end{equation*}}

\newcommand{\bel}{\begin{equation}\label}

\title{ Construction of the classical time crystal Lagrangians from Sisyphus dynamics and duality description with the
	Li\'{e}nard type equation
}
\author{
Partha Guha\footnote{E-mail: partha@bose.res.in; guha@ihes.fr}\\
IH\'ES, 35, Route de Chartres, \\
F-91440 Bures-sur-Yvette, France \\
\\
S.N. Bose National Centre for Basic Sciences \\
JD Block, Sector III, Salt Lake \\ Kolkata - 700106,  India \\
\and
A Ghose-Choudhury\footnote{E-mail aghosechoudhury@gmail.com}
\\
Department of Physics \\ Diamond Harbour Women's University,\\
D.H. Road, Sarisha, West-Bengal 743368, India\\
}

\date{ }

\maketitle

\begin{abstract}
We explore the connection between the  equations describing sisyphus dynamics and the generic Li\'{e}nard type or Li\'{e}nard II equation from the viewpoint of branched Hamiltonians. The former provides the appropriate setting for classical time crystal being derivable from a higher order Lagrangian. However it appears the equations of Sisyphus dynamics have a close relation with the Li\'{e}nard-II equation when expressed in terms of the `velocity' variable. Another interesting feature of the equations of Sisyphus dynamics  is the appearance of velocity dependent "mass function" in contrast to the more commonly encountered position dependent mass.  The consequences of such  mass functions seem to have connections to cosmological time crystals .
\end{abstract}

\smallskip

\noindent
{\bf Math Subject Classification Number:} 34C14, 34A34, 34C20.

\bigskip


\section{Introduction}
The issue of time independent classical dynamical systems
exhibiting motion in their lowest energy states has been
instrumental in the introduction of  a time analogue of spatial
order as in a crystalline substance \cite{SW1} (the so called
\textit{time crystals}) and its spontaneous breaking.
Generically time crystals refer to regular periodic behavior not in the spatial dimensions but in the time domain.
In the classical domain, time crystals are related to the periodic evolution of a system possessing the
lowest energy in which the motion does not
reach a standstill. Despite this apparent contradiction Shapere and Wilczek showed that if the
kinetic energy of a particle on a ring is a quartic function of its velocity
then the minimal energy state corresponds to a particle moving along a ring with a non-zero velocity.
In particular, the minimal time crystal Lagrangian for a single degree
of freedom is given by
$$ L = \frac{1}{12}\dot{y}^{4} - \frac{1}{2}\dot{y}^{2} - V(y), $$
where the higher temporal derivative terms violates time
translation symmetry. Note that Hamiltonian dynamics forbids the existence of classical time crystal,
since $H(p, y)$ minimizes at $ \frac{\partial H}{\partial p} = 0 = \frac{\partial H}{\partial y}$
with coordinate $y$ and conjugate momentum $p$. As a consequence in the minimum energy state $\dot{y}=\partial H/\partial p=0$ means that $y$ is a constant. However,
this negative result can be overcome if the structure of $H$ is such that the canonical momentum $p$ leads to a multivalued Hamiltonian as a function of $y$  with cusps at $\frac{\partial p}{\partial \dot{y}} = 0$, in which case
the Hamiltonian equations of motion cease to be valid. A common feature shared by all  the  models considered by Shapere and
Wilczek \cite{SW1,SW2} is that the energy function (Hamiltonian) or
Lagrangian systems  become multivalued in terms
of the canonical phase space variables.

\smallskip

The the study of  systems with
non-standard and/or non-convex Lagrangians especially with regard
to spontaneous breaking of time translation symmetry has also been investigated by a number of other authors following the initial work of Shapere and Wilczek \cite{ZYX, ZXY1, Zhao, Curtright}.\\

 Of late it has become clear that, for special kinds of mechanical
systems, there are choices of Hamiltonian structures in which certain fundamental aspects
of classical canonical Hamiltonian mechanics are changed. It has been
observed in \cite{ZYX,ZXY1,Zhao}, that one can change the phase space variables
which makes the Hamiltonian and symplectic structures on the phase space simultaneously
well defined at the price of introducing a non-canonical symplectic structure.
Curtright and Zachos \cite{Curtright}, on the other hand, studied certain simple unified Lagrangian
prototype systems which by virtue of non-convexity in their velocity dependence branch into
double-valued (but still self-adjoint) Hamiltonians.

\smallskip


 In this context it is
therefore natural to investigate the issue of time translation
breaking from the perspective of second-order differential
equations within the general framework of Lagrangian/Hamiltonian
mechanics and this is in fact constitutes our primary motivation. There are several interesting applications of time
crystal in molecular phenomena, ion trapping problem etc. \cite{DNPW,LGYQ}. The classical time crystal model
is a generalization of the Friedmann-Robertson-Walker (FRW) cosmology endowed with noncommutative geometry corrections
\cite{DPGP}.

\subsection{Motivation, result and organization}
As already mentioned in the introduction in \cite{SW1} Shapere and Wilczek showed  that a classical
system can reveal periodic motion in its lowest energy state. This essentially occurs when
the Hamiltonian is a multivalued function of the momentum with cusps
precisely corresponding to the  energy minima. The multivaluedness of the Hamiltonian is a consequence of the fact that in
theories containing higher powers of the velocities there are generally several solutions to the inversion equations between the velocities and the momenta. The latter also prevents the use of  the Ostragradsky method.  A particularly useful method for dealing with higher derivative theories is the Dirac formalism wherein one enlarges the phase space by including new additional variables while imposing suitable constraints. The advantage of the method is that it allows one to define  correct phase space coordinates in which one can solve the constraints and reduce the number of variables to the original number of degrees of freedom and thereby construct a Hamiltonian which is single valued with a canonical symplectic structure on the phase space. In this sense the problem of dealing with higher derivative theories and multivalued Hamiltonians may be dealt with within the canonical formalism. Our chief motivation is to understand the  duality between the Li\'{e}nard
  equation of the second kind  and the equation of motion of Sisyphus dynamics introduced by Shapere and Wilczek by invoking the Dirac formalism as formulated by Avraham and Brustein in \cite{AB}.\\

\noindent

In \cite{Sabatini} the author studied the second class of Li\'enard system
$\ddot{x} + f(x)\dot{x}^{2} + g(x) = 0$, with a center at the origin 0 and investigated
conditions under which it exhibited isochronicity.
The Li\'{e}nard equation of the second kind admits a Hamiltonian description using the Jacobi Last Multiplier
and it has a profound applications
in isochronous systems and the Hamiltonization of the Painlev\'e-Gambier type equations \cite{CG1,CG,GC}.
It should be noted that one finds
 up to a constant shift, the square of this Hamiltonian describes
systems giving rise to spontaneous time translation symmetry breaking provided the
potential function is negative \cite{GCPG2}. The phenomenon of spontaneous symmetry breaking is essentially connected to
Lagrangian theories with higher powers of velocities in the kinetic energy term. In this article by using the equations of Sisyphus
dynamics we investigate the  dual picture of symmetry breaking in terms
of the Li\'{e}nard  equation of the second kind with a Hamiltonian  involving a
position dependent mass and Lagrangian for a single degree
of freedom  given by $L = \frac{1}{4}\dot{y}^4 - \frac{1}{2}\dot{y}^2 - V(y)$.

It is quite easy to construct models of time-independent,
conservative dynamical systems with local ground states in
which $\phi_x \neq 0$. The potential energy
$$ V(\phi) = c_1 \phi_{x}^{4} - c_2\phi_{x}^{2}, $$
exhibits space translation is spontaneously broken in the ground states.
Our case is time-dependent one, where the potential energy
$$ \tilde{V}(\phi) = b_1 \phi_{t}^{4} - b_2\phi_{t}^{2}$$
shows spontaneous breaking of time translation
The condition $\phi_t \neq 0$ at the ground state seems to imply that
the system undergoes perpetual motion in its lowest energy.

\bigskip

The main result of the paper is given below.
\begin{prop}
Consider the system of equations describing Sisyphus dynamics
$$\mu\ddot{x}=f^\prime(x)\dot{y}-g^\prime(x)$$
$$\dot{x}f^\prime(x)=-V^\prime (y).$$
Define an invertible auxiliary function $h(x) $ such that
$$g(x)=\int f^\prime(x) h(x) dx.$$
\begin{enumerate}
\item In the limit as $\mu\rightarrow 0$ the equations of motion expressed in
terms of the $y$ coordinate has the Newtonian form
$$m(\dot{y}) \ddot{y}=-V^\prime(y)$$ with the mass function being
manifestly velocity dependent  and is given by
$$m(\dot{y})=h^{-1\prime}(\dot{y}) f^\prime(h^{-1}(\dot{y})).$$
In particular, for (1) $h(x) = x$, (2) $h(x) = x^n$,  (3) $h(x) = x^2 + a$ and (4) $h(x) = \frac{b-xd}{cx -a}$, \;\; $ad-bc\ne 0$
	we obtain the corresponding Lagrangians (1) $L_1 = \frac{\dot{y}^4}{12} - \frac{\dot{y}^2}{2} - V(y)$,
	(2) $L_2 = \frac{n}{3(n+3)}\dot{y}^{\frac{n+3}{n}}-\frac{n}{(n+1)}\dot{y}^{\frac{n+1}{n}}  - V(y)$,
	(3) $L_3 = \frac{2}{15}(\dot{y} -a)^{5/2} - \frac{2}{3}(\dot{y} -a)^{3/2} - V(y)$,
	(4) $L_4=\frac{1}{c}\left[\left(1-\frac{a^2}{c^2}\right)\frac{\Delta}{c}\log|c\dot{y} +d|-\frac{a\Delta^2}{c^3(c\dot{y} +d)}
+\frac{\Delta^3}{6c^3}\frac{1}{(c\dot{y} +d)^2}\right] -V(y)$ where $\Delta=ad-bc\ne 0$
respectively for the $y$ coordinate equations.
\item The equation of motion when expressed in terms of the $x$
coordinate  ( with the potential function taken to be $V(y)=y^2/2$ is of
the Lienard-II type and admits a Hamiltonian description.
\item The multivaluedness of the Hamiltonian is studied by employing
Dirac brackets to construct the appropriate Legendre transformation and the
resulting Hamilton's equations after a change of variable reproduces the
Lienard -II equation upon elimination of one of the dynamical variables.
This serves to illustrate the duality between the two descriptions.
\end{enumerate}

\end{prop}

The computation of the proof is given in Section 4.
Case (1), i.e., $h(x) = x$ coincides with the result of Shapere and Wilczek.

\bigskip

The {\bf Organization} of the article is as follows. In section 2 we briefly outline the Lagrangian and Hamiltonian features of
the Li\'{e}nard equation of the second kind. We then consider the equations for Sisyphus dynamics and establish its connection with  the Li\'{e}nard equation of the second kind in Section 3. In the next section we analyse the duality between the Li\'{e}nard equation and the equations of        Sisyphus dynamics when expressed in terms of the $x$ variable. We give the proof of our main result in section 4.1. Finally in Section 5 we outline an alternative procedure for arriving at the notion of spontaneous symmetry breaking.

\section{The Li\'{e}nard-II equation and its Hamiltonian aspects}

The generic form of the Li\'{e}nard
equation  of the second kind is
\be\label{2.4}\ddot{x}+f(x)\dot{x}^2 +g(x)=0.\ee  Using the concept of a Jacobi Last Multiplier (JLM) one can  derive a suitable Lagrangian for this equation \cite{GC}. The JLM for a second-order ordinary differential equation is defined to be  a solution of the equation
\be\label{JLMdef} \frac{d}{dt}\log M +\frac{\partial \mathcal{F}(x, \dot{x})}{\partial \dot{x}}=0, \;\;\;\mbox{where}\;\; \mathcal{F}(x, \dot{x})=-f(x)\dot{x}^2-g(x).\ee
In the present case the JLM  for (\ref{2.4}) is given by
\bel{2.5}M(x)=e^{2F(x)},\;\;\;\mbox{with}\;\;\;F(x):=\int^x f(s) ds.\ee  There exists a close  connection between the JLM and the Lagrangian which is provided by \cite{Whittaker}
\bel{Lagdef} M=\frac{\partial^2 L}{\partial\dot{x}^2}.\ee
In view of (\ref{2.5})
it follows  from (\ref{Lagdef}) that a Lagrangian  for the Li\'{e}nard-II equation (\ref{2.4}) is
\bel{2.6}
L(x,\dot{x})=\frac{1}{2}e^{2F(x)}\dot{x}^2-V(x),\ee where the
potential term \bel{2.7} V(x)=\int^xe^{2F(s)}g(s) ds.\ee Clearly
the conjugate momentum \bel{2.8} p:=\frac{\partial
L}{\partial\dot{x}}=\dot{x}e^{2F(x)}\;\;\mbox{implies}\;\;\dot{x}=pe^{-2F(x)},\ee
and the final expression for the Hamiltonian using the standard Legendre transformation yields:
\bel{2.9}H=\frac{p^2}{2M(x)}+\int^x M(s) g(s) ds,\ee where
$p=M(x)\dot{x}$ and $M(x)=\exp(2F(x))$ with $F(x)=\int^x f(s)ds$.
The canonical variables are $x$ and $p$ and they satisfy the standard
Poisson brackets  $\{x, p\}=1$. In terms of the canonical Poisson brackets the equations of motion appear as
 \bel{2.8a} \dot{x}=\{x, H\}=\frac{p}{M(x)}, \;\;\;\dot{p}=\{p, H\}=\frac{M^\prime(x)}{2M(x)}p^2-M(x)g(x),\ee
 from which we can recover (\ref{2.4}) upon elimination of the conjugate momentum $p$.
 Here we have purposely written the Hamiltonian $H$ in terms of
the last multiplier $M(x)$ to highlight the latter's role as a position
dependent mass term.

\bigskip

\noindent
As for the existence of a minima of $H$, considered as a function of $x$
and $p$, it is necessary that \be \frac{\partial H}{\partial
x}=0\;\;\;\mbox{and}\;\;\;\frac{\partial H}{\partial p}=0\ee whose
solutions then define the stationary points. The former yields
$$-p^2\frac{M^\prime(x)}{2M^2(x)}+M(x)g(x)=0$$ while the latter
implies $p/M(x)=0$. Therefore the stationary points are
characterized by $p=0$ and the value(s) of $x$ for which $g(x)=0$.
If $x=x^\star$ denotes a root of $g(x)=0$ then $(x^\star, p=0)$ is
a stationary point $(s.p)$. For a $s.p$ to be a minimum one
requires that the principal minors of
$$\Delta=\left| \begin{array}{ccc}H_{xx} & H_{xp}\\H_{px} &
H_{pp}\end{array}\right|_{s.p}$$ be positive definite, i.e.,
$$ g^\prime(x^\star)>0\;\;\;\mbox{ and }\;\;\;M(x^\star)g^\prime(x^\star)>0.$$
Consistency therefore requires  $M(x^\star)>0$. Note that
$M(x)$, which may be thought of as some kind of 'effective mass'
such as within a spatial crystal, may be negative for $x\ne
x^{\star}$. Clearly the fact that $p=0$ in the minimum energy
state (ground state) of the system precludes the possibility of
any motion. However as explained in Section 5 in order to arrive at spontaneous time translation symmetry breaking we need to consider a modification of the above Hamiltonian (\ref{2.9}).

\section{ Sisyphus dynamics}
The equations of motion for Sisyphus dynamics are given by
\be\label{Sisy1} \mu\ddot{x}=f^\prime(x) \dot{y}-g^\prime(x), \;\;\;\dot{x}f^\prime(x)=-V^\prime(y).\ee
In \cite{SW1} the authors consider the limit where the mass $\mu\rightarrow 0$. By making specific choices for the functions
\be f = \frac{1}{3}x^3 - x, \qquad g = \frac{1}{4}x^4 - \frac{1}{2}x^2, \ee
it immediately   from the first equation of the Sisyphus dynamics (\ref{Sisy1}) that as $\mu\rightarrow 0$, $ \dot{y} = x$ and
 the equation of motion for the $y$ coordinate turns out to be
 \be \label{yeqn} (\dot{y}^{2} - 1)\ddot{y} = -V^{\prime}(y), \qquad \hbox{ where } \qquad f^{\prime} = \dot{y}^{2} - 1.
\ee
on using the
second equation $\dot{x} = - \frac{V^{\prime}(y)}{f^{\prime}}$.

\smallskip

\begin{center}
\includegraphics[width=3.0 in]{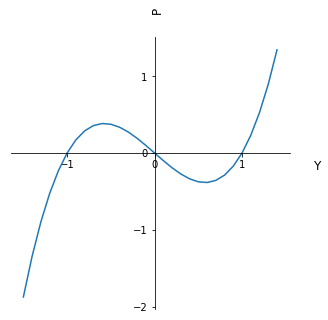}
\begin{figure}[h]
\caption{Plot of $p = y^3 -y$}
\end{figure}
\end{center}

\bigskip

\begin{center}
\includegraphics[width=3.0 in]{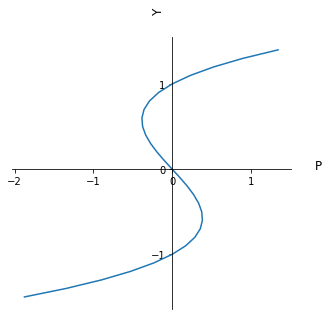}
\begin{figure}[h]
\caption{Plot of the inverse function \big($-\frac{(\sqrt{3}(\sqrt{27x^2 - 4} - 9x)^{1/3}}{2^{1/3}3^{2/3}}$ - conjugate \big)
given in fig.1 }
\end{figure}
\end{center}

\smallskip

This equation is clearly of the Newtonian type with `mass' being velocity dependent. Such equations are less common than the more familiar equations with position dependent mass (PDM)
in nonlinear dynamics. Thus (\ref{yeqn}) equation is really to be viewed as a  velocity dependent mass (VDM) equation of the Newtonian type.

\smallskip

\noindent
Let us now focus on the equation for the variable $x$. To this end we
 begin by assuming that the function $V(y)$ is single valued with an isolated minimum at $y=0$. We shall assume for concreteness, $V(y)=y^2/2$, so that the latter equation in (\ref{Sisy1}) leads to $y=-\dot{x}f^\prime(x)$. Eliminating now $\dot{y}$ from the first equation we arrive at the following second-order differential equation
\be\label{L2}\ddot{x}+\frac{f^\prime f^{\prime\prime}}{\mu + f^{\prime 2}}\dot{x}^2+\frac{g^\prime (x)}{\mu + f^{\prime 2}}=0.\ee
Clearly  this is an equation of the Li\'{e}nard-II type and in the limit as $\mu\rightarrow 0$ we have
$$\ddot{x}+\frac{f^{\prime\prime}}{f^\prime}\dot{x}^2+\frac{g^\prime}{f^{\prime 2}}=0,$$
which still retains the form of the Li\'{e}nard-II equation.

\noindent
 Let us for the time being dispense with the explicit forms of the functions $f$ and $g$ and  go back to (\ref{Sisy1}) and consider the limit $\mu\rightarrow 0$; then we have $\dot{y}=g^\prime/f^\prime=h(x)$, (say) and $\dot{x}f^\prime(x)=-V^\prime(y)$. In this scenario we may derive a corresponding second-order equation for the variable $y$ by assuming $h(x)$ is invertible. We have  formally $\dot{x}=h^{-1 \prime}(\dot{y})\ddot{y}$ so that
\bel{3.5} h^{-1 \prime}(\dot{y})f^\prime(h^{-1}(\dot{y}))\ddot{y}=-V^\prime(y),\ee which may be regarded as a generalized version of (\ref{yeqn}.) Clearly (\ref{3.5}) is
a Newtonian equation of motion  and it is plain that the mass function is given by
$$m(\dot{y})=h^{-1 \prime}(\dot{y})f^\prime(h^{-1}(\dot{y})),$$ which is velocity dependent.
As the JLM is related to the Lagrangian of a system by the formula, $M={\partial^2 L}/{\partial \dot{y}^2}$,  and in one dimension represents physically the mass function of the system, therefore for the inverse problem we are justified in assuming
\bel{3.5a}\frac{\partial^2 L}{\partial \dot{y}^2}=h^{-1 \prime}(\dot{y})f^\prime(h^{-1}(\dot{y})),\ee which implies
that the Lagrangian is of the form
\bel{3.5b} L=\int d\dot{y}\int^{\dot{y}} dz h^{-1 \prime}(z)f^\prime(h^{-1}(z)) -V(y).\ee
It will be noticed that  if the functional forms of $f$ and $g$ are chosen as in \cite{SW3} then their substitution into the above formula (\ref{3.5b}) for the Lagrangian gives
\bel{3.5c}L=\frac{1}{12}\dot{y}^4-\frac{1}{2}\dot{y}^2-V(y),\ee as it should. The velocity dependent mass functions in this case is $M(\dot{y})=(\dot{y}^2-1)$. If we assume the potential function $V(y)=y^2/2$ then the conjugate momenta turns out to be $p=\dot{y}^3-\dot{y}$ and from $H=p\dot{y}-L$ we obtain
$$H=\frac{11}{12}\dot{y}^4-\frac{1}{2}\dot{y}^2+\frac{1}{2}y^2.$$ Obviously the Hamiltonian here is not written in the correct variables. To express the Hamiltonian  in terms of the canonical variables it is necessary to invert the equation $p=\dot{y}^3-\dot{y}$. This clearly leads to multi valued solutions of $\dot{y}$ in terms of $p$ as depicted in the figures shown below.
Therefore the  Hamiltonian with $\dot{y}^{4}$ like terms cannot be expressed as a single valued function of the phase space variables. The energy has to be regarded as a multivalued function of the momentum $p$ with cusps
$\frac{\partial p}{\partial \dot{y}} = 0$.
At the cusps the usual condition that the gradient should vanish at
a minimum does not apply and hence the conjugate momentum is not a good variable for writing down the corresponding Hamiltonian uniquely.

One can of course propose a more general form of the mass function/JLM namely
$M=a_0(y)\dot{y}^2+a_2(y)$. The resulting Lagrangian  is then given by
$$ L=\frac{a_0(y)}{12}\dot{y}^4+\frac{a_2(y)}{2}\dot{y}^2 -V(y)$$ and such a Lagrangian has indeed been considered in
\cite{DPGP} in the context of cosmological time crystals in quadratic gravity.

When the above Lagrangian is inserted into the the Euler-Lagrange equation then it yields the following equation,
\bel{3.5d}(a_0(y)\dot{y}^2+a_2(y))\ddot{y}+\frac{1}{4}a_0^\prime(y)\dot{y}^4 + \frac{1}{2}a_2^\prime(y)\dot{y}^4 +B^\prime(y)=0.\ee
This is a higher degree version of the usual Li\'{e}nard-II equation. Clearly when $a_0$ and $a_2$ are constants we get
$$(c\dot{y}^2 +a)\ddot{y} + B^\prime(y)=0,$$
which is nothing but the (\ref{yeqn}).

\smallskip

We have already seen that  when, $V(y)=y^2/2$, then upon eliminating $\dot{y}$ from the Sisyphus dynamics we obtain the equation (3.4) which we write as
$$ \ddot{x} + \tilde{f}(x)\dot{x}^{2} + \tilde{g}(x) = 0, \quad \hbox{ with } \,\,\,
\tilde{f}(x) = \frac{f^\prime f^{\prime\prime}}{\mu + f^{\prime 2}}, \,\,\,\,
\tilde{g}(x) = \frac{g^\prime (x)}{\mu + f^{\prime 2}}. $$
The solution of the JLM is given by $$ M_{Sisy} = e^{2\tilde{F(x)}} =
(\mu + f^{\prime 2}), \qquad \tilde{F}(x) := \int^x \tilde{f}(s)ds, $$
and the Lagrangian of the equation is
\be
L_{Sisy}(x,\dot{x}) = \frac{1}{2}(\mu + f^{\prime 2})\dot{x}^{2} - V(x),
\ee
where
\be
V(x) = \int^x (\mu + f^{\prime 2}(s))\tilde{g}(s)\, ds = g(x).
\ee
Thus the momentum is given by $ p = \dot{x}(\mu + f^{\prime 2}(x))$ and the corresponding Hamiltonian is
\be
H_{Sisy}(x,p) = \frac{p^2}{2(\mu + f^{\prime 2}(x))} + g(x).
\ee

In this article we retain the general forms of $f(x)$ and $g(x)$ for the most part and attempt to  understand the duality between the  velocity dependent mass form of the Sisyphus equation of motion in terms of the $y$ variable and the corresponding $x$ equation of motion which is of the Li\"{e}nard-II type.


\section{Duality from Dirac brackets and Legendre transformation}

We consider again the  expression for the Lagrangian as given by (\ref{3.5b}), i.e,
\be \label{Lagsisy}
L=\int d\dot{y}\int^{\dot{y}} dz h^{-1 \prime}(z)f^\prime(h^{-1}(z)) -V(y).
\ee
As already illustrated higher powers of $\dot{y}$ in the R.H.S. of (\ref{Lagsisy}) lead to a complicated
algebraic equation connecting the canonical momentum
\be p_y = \frac{\partial L}{\partial \dot{y}} = \int^{\dot{y}} dz h^{-1 \prime}(z)f^\prime(h^{-1}(z)) \ee
and the velocity $\dot{y}$. This usually leads to a multivalued Hamiltonian
in terms of $p_y$.

\smallskip

Following Avraham and Brustein \cite{AB} we treat the velocity $\dot{y}$ as a new coordinate $Q$ and after introducing a Lagrange
multiplier write a modified Lagrangian
\be \tilde{L} = L(y,Q) + \lambda (\dot{y} - Q). \ee
A standard Legendre transformation leads to the Hamiltonian
\be \tilde{H} = \dot{y}\frac{\partial \tilde{L}}{\partial \dot{y}} - \tilde{L} = \lambda Q - L(y,Q). \ee
The primary or \textit{first class }constraints are given by
\be
\phi_1 = p_Q = \frac{\partial \tilde{L}}{\partial Q} = 0, \quad \phi_3 = p_3 - \lambda \approx 0, \quad
\phi_4 = p_{\lambda} \approx 0, \ee
where $p_y = \frac{\partial \tilde{L}}{\partial \dot{y}} = \lambda$.
As the constraints $\phi_3$ and $\phi_4$ are not dynamical we can insert their values in the Hamiltonian
and write
\be
H_1 = \tilde{H} + \mu_1\phi_1 + \mu_3\phi_3 + \mu_4\phi_4 = \lambda Q - L(y,Q) + \mu_1p_Q.
\ee
The secondary or \textit{second class }constraint follows from the condition
$$
\dot{\phi} = \{ \phi_1, H_1 \} = 0 \Rightarrow \phi_2 = \frac{\partial L}{\partial Q} - \lambda \approx 0,
$$
using $\{Q,p_Q \} = 1$. Hence we have the total Hamiltonian
\be
H_T = H_1 + \mu_2\phi_2 = \lambda Q - L(y,Q) + \mu_1p_Q + \mu_2(\frac{\partial L}{\partial Q} - \lambda).
\ee
To construct Dirac brackets we note that
\be \{\phi_1, \phi_2 \} = - \frac{\partial^2 L}{\partial Q^2}, \ee and the Dirac matrix
$C$ is defined as a skew symmetric matrix having entries $C_{ij}=\{\phi_i, \phi_j\}$ so that in the present case  has the appearance
 $$ C = - \frac{\partial^2 L}{\partial Q^2}\left(\begin{array}{cc}
                       0 & 1 \\
                                           -1  & 0  \\
                                    \end{array}\right), $$
 with its  inverse given by
 $$C^{-1}=\frac{1}{\frac{\partial^2 L}{\partial Q^2}}\left(\begin{array}{cc}
                       0 & 1 \\
                                           -1  & 0  \\
                                    \end{array}\right). $$
The expression for the Dirac bracket of $A$ and $B$ is
\be
\{A,B \}_D = \{A,B\} - \frac{1}{\frac{\partial^2 L}{\partial Q^2}}\big(\{A,\phi_1\}\{\phi_2,B\} - \{A,\phi_2\}\{\phi_1,B\} \big).
\ee
On shell, after inserting the constraints the Hamiltonian becomes
\be
H_F = Q \frac{\partial L}{\partial Q} - L(y,Q). \ee
However the Dirac bracket of the phase space variables $(y,Q)$ is not unity but
\be
\{y,Q\}_D = -\frac{1}{\frac{\partial^2 L}{\partial Q^2}}. \ee
To arrive at a canonical structure one is therefore led to consider a change of variables
\be
(y,Q) \longrightarrow (F(y,Q), G(y,Q), \quad \hbox{ s.t. } \quad \{F,G\}_D =1,
\ee
  which translates to the requirement
\be
\frac{\partial F}{\partial y}\frac{\partial G}{\partial Q} - \frac{\partial F}{\partial Q}\frac{\partial G}{\partial y} = \frac{\partial^2 L}{\partial Q^2}
= {h^{-1}}^{\prime}(Q)f^{\prime}(h^{-1}(Q)). \ee
The simple choice $G = Q$ then implies $F = y{h^{-1}}^{\prime}(Q)f^{\prime}(h^{-1}(Q))$ and we may therefore  re-express $H_F(y,Q)$ in terms of variables
$F$ and $G$. As a consequence we have
\bel{can1}
\dot{F}(y,Q) = \{(F(y,Q), H(F(y,Q), G(y,Q)) \}_D = \{F,F\}_D\frac{\partial H}{\partial F} +
\{F,G\}_D\frac{\partial H}{\partial G} = \frac{\partial H}{\partial G},
\ee and similarly
\bel{can2}
\dot{G}(y,Q) = \{(G(y,Q), H(F(y,Q), G(y,Q)) \}_D = -\frac{\partial H}{\partial F}.
\ee
Thus in summary  we have the following change of variables
\be
Q = G, \quad y = \frac{F}{{h^{-1}}^{\prime}(Q)f^{\prime}(h^{-1}(Q))},
\ee

together with the Hamiltonian written in terms of the new variables $(F, G)$ which leads to a canonical structure:
\bel{canHam}
H(F,G) = G\int^G dz {h^{-1}}^{\prime}(z)f^{\prime}(h^{-1}(z))  - \int dG \int^G dz {h^{-1}}^{\prime}(z)f^{\prime}(h^{-1}(z))
+ V\left(\frac{F}{{h^{-1}}^{\prime}(G)f^{\prime}(h^{-1}(G)} \right).
\ee
Using (\ref{can1}) and (\ref{can2}) the canonical equations of motion are

\be
\dot{F} = \int^a dz {h^{-1}}^{\prime}(z)f^{\prime}(h^{-1}(z)) - \int^G dz {h^{-1}}^{\prime}(z)f^{\prime}(h^{-1}(z))
+ V^{\prime}(r)F \frac{\partial}{\partial G}\Big(\frac{1}{{h^{-1}}^{\prime}(G)f^{\prime}(h^{-1}(G))} \Big)
\ee
\be\dot{G} = - \frac{\partial H}{\partial F} = -\frac{V^{\prime}(r)}{{h^{-1}}^{\prime}(G)f^{\prime}(h^{-1}(G))}, \;\;\;
\mbox{with}\;\;r=\frac{F}{{h^{-1}}^{\prime}(G)f^{\prime}(h^{-1}(G))}.
\ee

\subsection{Some cases of invertible functions $h^{-1}(z)$ }

We  consider the following cases.
\begin{enumerate}
\item $h^{-1}(z) = z$ or $h(z) = z$.
\item $h^{-1}(z) = z^{1/n}$, i.e. $h(z) = z^n$.
\item  $h^{-1}(z) = \sqrt{z-a}$, $ a > 0$, $z \geq a$, i.e., $h(z) = z^2 + a$.
\item $h^{-1}(z) = \frac{az+b}{cz+d}$, then $h(z) = \frac{b-zd}{cz -a}$, \;\; $ad-bc\ne 0$.
\end{enumerate}
Note that as $h(z) = \frac{g^{\prime}(z)}{f^{\prime}(z)}$,  once $h(z)$ and $f(z)$ have been chosen
then $g(z)$ is obtained from
$$ g(z) = \int {f^{\prime}(z)}{h(z)} dz. $$
The Lagrangians corresponding to the above cases with the explicit choice $f(z)=z^3/3-z$ are given below:
\begin{enumerate}
\item $L_1 = \frac{\dot{y}^4}{12} - \frac{\dot{y}^2}{2} - V(y)$.
\item $L_2 = \frac{n}{3(n+3)}\dot{y}^{\frac{n+3}{n}}-\frac{n}{(n+1)}\dot{y}^{\frac{n+1}{n}}  - V(y)$.
\item $L_3 = \frac{2}{15}(\dot{y} -a)^{5/2} - \frac{2}{3}(\dot{y} -a)^{3/2} - V(y)$.
\item $L_4=\frac{1}{c}\left[\left(1-\frac{a^2}{c^2}\right)\frac{\Delta}{c}\log|c\dot{y} +d|-\frac{a\Delta^2}{c^3(c\dot{y} +d)}
+\frac{\Delta^3}{6c^3}\frac{1}{(c\dot{y} +d)^2}\right] -V(y)$ where $\Delta=ad-bc\ne 0$
\end{enumerate}

{\bf Case 1 :} In this case it follows that $Q = G$, $y = \frac{F}{(G^2 -1)}$,  if $V(z)$ is taken as $\frac{1}{2}z^2$. Hence from the Hamiltonian
\be
H = G \int^G dz (z^2 -1) - \int dG \int^G dz(z^2 -1) + V(\frac{F}{G^2 -1 })
\ee
we obtain
$$ H = \frac{1}{4}G^4 - \frac{1}{2}G^2 + \frac{1}{2}\frac{F^2}{(G^2 -1)^2}. $$
This yields
\bel{xx}
\dot{F} = \frac{\partial H}{\partial G} = G^3 - G^2 - \frac{2F^2G}{(G^2 -1)^3}, \qquad \dot{G} = -\frac{\partial H}{\partial F}
= - \frac{F}{(G^2 -1)^2}.
\ee
If we switch back to the old variable then  as,
$\dot{y} = Q = G$,  the second equation in (\ref{xx}) actually coincides with the result of Shapere and Wilczek \cite{SW3}, namely $(\dot{y}^{2} - 1)\ddot{y} = -y$, as already mentioned.\\

\noindent
{\bf Case 2:} In this case  it follows that
$$H(F, G)=\frac{1}{n+3}G^{\frac{n+3}{n}}-\frac{1}{n+1}G^{\frac{n+1}{n}} +V(\frac{F}{s})$$ where
$$s=\frac{1}{n}\left(G^{\frac{3}{n}-1}-G^{\frac{1}{n}-1}\right)$$
This leads to the canonical equations of motion, namely
\be \dot{F}=\frac{\partial H}{\partial G}=\frac{1}{n}\left(G^{\frac{3}{n}}-G^{\frac{1}{n}}\right)-V^\prime(\frac{F}{s})\frac{F}{ns^2}
\left((\frac{3}{n}-1)G^{\frac{3}{n}-2}-(\frac{1}{n}-1)G^{\frac{1}{n}-2}\right),\ee
\be \dot{G}=-\frac{\partial H}{\partial F}=-V^\prime(\frac{F}{s})\frac{1}{s}.\ee
 If one assumes $V(z)=z^2/2$ then  straightforward calculation gives the following forms of the equations of motion
 \be \dot{F}=\frac{1}{n}\left(G^{\frac{3}{n}}-G^{\frac{1}{n}}\right)-
 \frac{F^2}{ns^3}\left((\frac{3}{n}-1)G^{\frac{3}{n}-2}-(\frac{1}{n}-1)G^{\frac{1}{n}-2}\right),\ee
\be \dot{G}=-\frac{F}{s^2}.\ee
Upon eliminating $F$ from the above system we then arrive at the second-order equation
\be \ddot{G}+\frac{\dot{G}^2}{ns}\left((\frac{3}{n}-1)G^{\frac{3}{n}-2}-(\frac{1}{n}-1)G^{\frac{1}{n}-2}\right)+
\frac{1}{ns^2}\left(G^{\frac{3}{n}}-G^{\frac{1}{n}}\right)=0.\ee
This is an equation of the form of a Li\'{e}nard equation of the second kind.\\

\noindent
{\bf Case 3:} This is a special case of the situation considered above, involving a translation of the $z$ coordinate after setting the value of the exponent $n=2$ and therefore we do not consider  its details.\\

\noindent
{\bf Case 4:}  In this case the Lagrangian is given by
$$L=\frac{1}{c}\left[\left(1-\frac{a^2}{c^2}\right)\frac{\Delta}{c}\log|c\dot{y} +d|-\frac{a\Delta^2}{c^3(c\dot{y} +d)}
+\frac{\Delta^3}{6c^3}\frac{1}{(c\dot{y} +d)^2}\right] -V\left(\frac{F}{s}\right)$$
where
$$s(\dot{y})=\Delta\left(\frac{(a\dot{y} +b)^2}{(c\dot{y} +d)^4}-\frac{1}{(c\dot{y} +d)^2}\right), \;\;\mbox{with}\;\;\Delta=ad-bc\ne 0$$
 In this case the Hamiltonian  when written in terms of the transformed variables $(F, G)$ is given by
 $$H(F, G)=\frac{\Lambda_1G+\Lambda_2}{(cG+d)}+\frac{\Lambda_3G+\Lambda_4}{(cG+d)^2}+\Lambda_5\log|cG+d| +V\left(\frac{F}{u(G)}\right)$$
 where the coefficients have the following values
 $$\Lambda_1=\frac{\Delta}{c}\left(1-\frac{a^2}{c^2}\right), \;\;\;\Lambda_2=\frac{a\Delta^2}{c^4},\;\; \Lambda_3=\frac{a\Delta^2}{c^3},\;\;\;\Lambda_4=-\frac{\Delta^3}{6c^4}, \;\;\Lambda_5=-\frac{\Lambda_1}{c}$$
Assuming as before $V(z)=z^2/2$ the equations of motion following from the above Hamiltonian are given by
$$\dot{F}=\frac{\partial H}{\partial G}=\frac{c\Lambda_5}{(cG +d)}+\frac{\Lambda_1 d-\Lambda_2 c+\Lambda_3}{(cG +d)^2}
-\frac{2c(\Lambda_3 G +\Lambda_4)}{(cG +d)^3}-\frac{s^\prime(G)}{s^3(G)} F^2,$$
$$\dot{G}=-\frac{F}{s^2(G)}.$$  Upon elimination of $F$ we arrive at the following second-order equation of the Li\'{e}nard type namely:
$$\ddot{G}+\frac{s^\prime(G)}{s(G)}\dot{G}^2+K(G)=0,$$
where
$$K(G)=\frac{1}{s^2(G)}\left[\frac{c\Lambda_5}{(cG +d)}+\frac{\Lambda_1 d-\Lambda_2 c+\Lambda_3}{(cG +d)^2}
-\frac{2c(\Lambda_3 G +\Lambda_4)}{(cG +d)^3}\right],$$ and
$$s(G)=\Delta\left[\frac{(aG+d)^2}{(cG+d)^4}-\frac{1}{(cG+d)^2}\right].$$

\section{A modified Hamiltonian  and spontaneously broken time
translation symmetry}

Consider a one-dimensional generalized Hamiltonian system $\widetilde{H} = {\cal F}(H)$
with Hamiltonian vector field given in terms of the canonical form
$$
{\Bbb X}_{\widetilde{H}} = \frac{\partial {\widetilde{H}}}{\partial p}\frac{\partial}{\partial x} -
\frac{\partial {\widetilde{H}}}{\partial x}\frac{\partial}{\partial p}, \qquad \{G,{\widetilde{H}}\} = \dot{G}.
$$
In the symplectic coordinates $(x,p)$ this is equivalent to canonical Hamiltonian equations
$$ \dot{x} = {\cal F}(H)^{\prime}\{x,H\}, \qquad \dot{p} = {\cal F}(H)^{\prime}\{p,H\}, \qquad \hbox{ where }\,\,\,\,   {\cal F}(H)^{\prime} > 0. $$
It may be easily verified that the above set of Hamiltonian equations may be obtained from the modified symplectic form
$\omega =  {\cal F}(H)^{\prime} dx \wedge dp$. Moreover this change of Hamiltonian structure
will not change the partition function, hence all thermodynamic quantities will remain unchanged.

\noindent
 Let us  consider  a new Hamiltonian \cite{ZXY1} defined by the square of $H$ as given by (\ref{2.9}) and a shift, i.e.,
\be\label{H2} \widetilde{H}=\left(\frac{p^2}{2M(x)}+\int^x M(s)
g(s) ds\right)^2+E_0=H^2+E_0,\ee where $E_0$ is an arbitrary constant. As the new Hamiltonian is anticipated to generate a dynamics which is distinct from that of $H$, let us also introduce the following Poisson structure
 $\{x, p\}=\xi(x,p)$ so that the equations of motion which follow from
\be\label{E1}
\dot{x}=\{x,\widetilde{H}\},\;\;\;\dot{p}=\{p,\widetilde{H}\}\ee
 give \be\dot{x}=2\xi H\frac{p}{M(x)}\ee \be\dot{p}= -2\xi
H\left(-\frac{M^\prime(x)}{2M^2(x)}p^2+M(x) g(x)\right).\ee At this point we need to make a clear distinction regarding the two Poisson structures we have introduced. It will be noticed that if one assumes $\{x, p\}=\xi(x, p)=\frac{1}{2H(x,p)} $ then we get back the original Li\'{e}nard-II equation (\ref{2.4}), if however we persist
with $\xi=1$, i.e., assume $x$ and $p$ are canonical then the
equation of motion resulting from the Hamiltonian $\widetilde{H}$
is of the form \be\label{E3}\ddot{x}+2H(f(x)\dot{x}^2+g(x))=0.\ee
Although (\ref{E3}) appears to be different from (\ref{2.4}) it is interesting to note that (\ref{E3})  can be mapped to the original set of
 Hamiltonian equations (\ref{2.8a}) by using a (nonlocal) Sundman transformation \cite{Sundman} through a transformation of the independent temporal variable $t$ to a
new independent variable $s$ given by $ds = 2H dt$, whence we obtain
\be {x}^{\prime}= \frac{p}{M(x)}, \qquad {p}^{\prime} = -\left(-\frac{M^\prime(x)}{2M^2(x)}p^2+M(x) g(x)\right),\ee
where $^\prime = \frac{d}{ds}$. In fact such transformations were used by Sundman while attempting to solve the restricted  three body problem.

\smallskip

As for the stationary points of the  Hamiltonian $\widetilde{H}$, these follow
from the solutions  of $\partial\widetilde{H}/\partial x=0$ and
$\partial\widetilde{H}/\partial p=0$. The latter yields either
$p=0$ or $H=0$. If $p=0$ then the former condition gives either
$H=0$ or $g(x)=0$, i.e $x=x^\star$. The pair $(x^\star, p=0)$
leads by the previous analysis to the case \be\label{E4}
\widetilde{H}_{min}=\left(\int^{x^\star} M(s) g(s)
ds\right)^2+E_0.\ee
From the above equation it is clear that the local minimum of $\widetilde{H}$ is in general greater than
the constant $E_0$ because the potential $V(x^\star)$ is not required to vanish at $x=x^\star$. As the stationary point corresponds to $p=0$ the time translation symmetry is not broken and we have the same situation as previously discussed in section 2.

However one also has now the possibility
wherein $H=0$ which implies that the locus of the stationary points lie on the
curve \cite{GCPG2}
\be\label{E3.7a}\frac{p^2}{2M(x)}+\int^x M(s) g(s) ds=0.\ee This condition
obviously implies that $\widetilde{H}$ has a minima with
$\widetilde{H}_{min}=E_0$ which is  less than that given by
(\ref{E4}). Now for real values of $p$ it is then necessary that
$$V(x)=\int^x M(s) g(s) ds<0.$$
The force $dV/dx$ is clearly not necessarily zero and motion can therefore occur in the ground state.
The existence of motion under such
circumstances is indicative of the spontaneous breaking of the
time-translation symmetry \cite{SW1}. This then provides an alternative procedure for obtaining motion in the minimal energy state.

\smallskip

\section{Conclusion}
The equations governing Sisyphus dynamics are expressed in terms of
 the auxiliary variable $x$ instead of $y$ as done in  \cite{SW3} by Shapere and Wilczek who assumed that the constitute functions $f$ and $g$ are such that $\dot{y}$,
 is linear in $\dot{y}$. In this paper we have considered a more general form such that
$\dot{y} = h(x)$, hence it is a function of the momentum conjugate to $y$.
In this situation we have derived a corresponding second-order equation,
for the `` momentum'' coordinate $x$, $\ddot{x} + f(x)\dot{x} + g(x) = 0$,
which belongs to the Li\'enard-II type.

\smallskip

In our computation we have considered $ h(x) = \frac{g^{\prime}(x)}{f^{\prime}(x)}$,
 for fixed $f(x) = x^3/3 - x$ and for different invertible functions $h(x)$ using which
we have obtained the corresponding classical time crystal Lagrangians.
The Lagrangians contain higher powers of the time derivative and display a nonlinear relationship
between the velocities and momenta, thus leading to a  multivalued canonical description.
The multivalued nature of phase space originates due to an inappropriate choice of coordinates.
Following Avraham and Brustein \cite{AB}, the Dirac formalism is introduced to define a generalized Legendre transformation.
In this paper we have shown that the phase space coordinates may be so chosen as to enable one to define a single valued Hamiltonian for the generalized  classical time crystal Lagrangians.

\smallskip

We have also outlined the the branched  Hamiltonian aspects of the Li\'enard type equation
corresponding to the `` momentum '' coordinate.
Although a number of articles have appeared on branched Hamiltonians, there appears to be no uniform
consensus on the physical interpretations of the results of these analyses.  The Hamiltonians
studied have almost invariably time independent. In this context it is interesting to mention that
there are examples of time-dependent Hamiltonian systems for which one can define a suitable conjugate
set of the canonical Hamilton's equations and they offer a alternative scenario to test for
multi-valuedness and branching to the Hamiltonian thereby leading to possibly a new dynamics.

\section*{ Acknowledgement}
We are grateful to Dr. Supriyo Datta for drawing the plots.
PG would like to express his gratitude to the members of IHES for their warm hospitality
where the final parts of the work has been done.

\end{document}